\documentclass[journal=nalefd,manuscript=letter]{achemso}

\usepackage[version=3]{mhchem} % Formula subscripts using \ce{}

\author{Yuhui He}
\affiliation[Chinese Academy of Sciences]
{Laboratory of Nano-Fabrication and Novel Devices Integrated Technology, Institute of Microelectronics, Chinese Academy of Sciences, Beijing 100029, China}
\author{Ralph H. Scheicher}
\email{ralph.scheicher@fysik.uu.se}
\author{Anton Grigoriev}
\affiliation[Uppsala University]
{Condensed Matter Theory Group, Department of Physics and Astronomy, Box 516, Uppsala University, SE-751 20 Uppsala, Sweden}
\author{Rajeev Ahuja}
\affiliation[Uppsala University]
{Condensed Matter Theory Group, Department of Physics and Astronomy, Box 516, Uppsala University, SE-751 20 Uppsala, Sweden}
\alsoaffiliation[Royal Institute of Technology]
{Applied Materials Physics, Department of Materials Science and Engineering, Royal Institute of Technology (KTH), SE-100 44 Stockholm, Sweden}
\author{Shibing Long}
\author{ZongLiang Huo}
\author{Ming Liu}
\email{liuming@ime.ac.cn}
\affiliation[Chinese Academy of Sciences]
{Laboratory of Nano-Fabrication and Novel Devices Integrated Technology, Institute of Microelectronics, Chinese Academy of Sciences, Beijing 100029, China}

\title[Edge-hydrogenated graphene for DNA sequencing]
{Enhanced DNA sequencing performance through edge-hydrogenation of graphene electrodes}

\begin{document}
%%%%%%%%%%%%%%%%%%%%%%%%%%%%%%%%%%%%%%%%%%%%%%%%%%%%%%%%%%%%%%%%%%%%%
%% The manuscript does not need to include \maketitle, which is
%% executed automatically.  The document should begin with an
%% abstract, if appropriate.  If one is given and should not be, the
%% contents will be gobbled.
%%%%%%%%%%%%%%%%%%%%%%%%%%%%%%%%%%%%%%%%%%%%%%%%%%%%%%%%%%%%%%%%%%%%%
\begin{abstract}
We propose using graphene electrodes with hydrogenated edges for solid-state nanopore-based DNA sequencing, and perform molecular dynamics simulations in conjunction with electronic transport calculations to explore the potential merits of this idea. The results of our investigation show that, compared to the unhydrogenated system, edge-hydrogenated graphene electrodes facilitate the temporary formation of H-bonds with suitable atomic sites in the translocating DNA molecule. As a consequence, the average conductivity is drastically raised by about 3 orders of magnitude while exhibiting significantly reduced statistical variance. We have furthermore investigated how these results are affected when the distance between opposing electrodes is varied and have identified two regimes: for narrow electrode separation, the mere hindrance due to the presence of protruding hydrogen atoms in the nanopore is deemed more important, while for wider electrode separation, the formation of H-bonds becomes the dominant effect. Based on these findings, we conclude that hydrogenation of graphene electrode edges represents a promising approach to reduce the translocation speed of DNA through the nanopore and substantially improve the accuracy of the measurement process for whole-genome sequencing.
\end{abstract}


Tremendous recent advances have been made in the fabrication of solid-state nanopores \cite{Wu2009,Taniguchi2009} and in their envisioned application for rapid whole-genome sequencing \cite{Storm2005,Zwolak2005,Lagerqvist2006,Iqbal2007,Dekker2007}. The basic concept centers around the idea that the four types of nucleobases occurring in DNA (adenine, thymine, cytosine, guanine; in the following abbreviated as A, T, C, G) possess different local electronic densities of states, which are electrically distinguishable and could thus in principle be used to differentiate between them. Very recently, the electrical detection of single isolated nucleotides residing between nanoelectrodes has been realized, identifying three of the four nucleotides based on a statistical distribution of electrical conductivity curves \cite{Tsutsui2010}. However, the development of solid-state nanopore-based DNA sequencing continues to struggle with a series of extremely challenging requisites, in particular single-base resolution during the polynucleotide translocation through the nanopore, optimized contrast in the electrical signals between the four different types of nucleotides, and a general improvement of signal-to-noise ratio \cite{Branton2008,Zwolak2008}.

\begin{figure*}[ht]
\includegraphics[width=16cm]{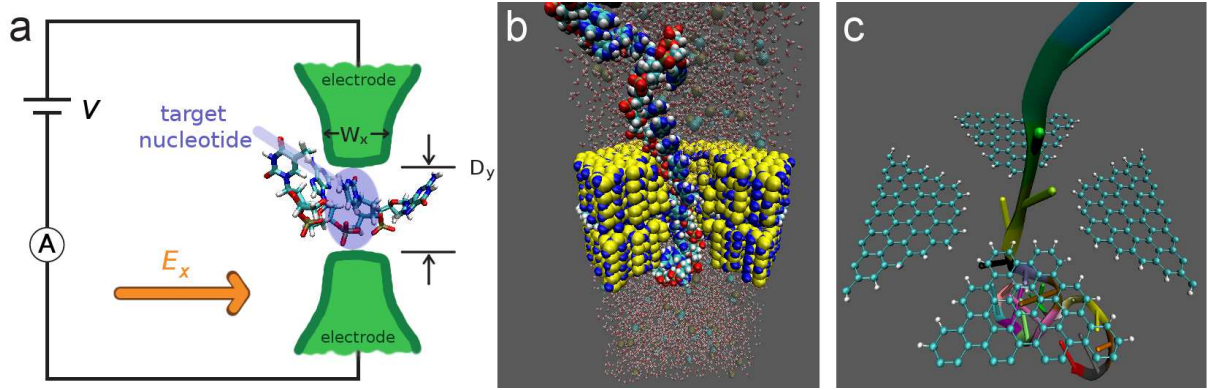}
\caption{(a) Schematic view of single-stranded DNA translocating through a nanopore under the application of a longitudinal electrical field $E_x$, while the transverse tunneling current is recorded for the purpose of sequencing. $W_{x}$ is the width of the transverse nanoelectrodes. $D_{y}$ is the inner diameter of the nanopore which characterizes the gap distance between opposing transverse nanoelectrodes. (b) Cross-section visualization (cut along the $x-z$ plane) of the complete atomistic setup employed in the present simulation work, showing the silicon-nitride membrane/nanopore (blue and yellow), a translocating single-stranded DNA molecule, as well as water molecules (red and white) and counter ions (ochre and cyan). (c) Cross-section visualization (cut along the $y-z$ plane) of the setup, showing only the edge-hydrogenated graphene electrodes (cyan and white) and a cartoon-version of the translocating single-stranded DNA molecule with the sugar-phosphate backbone represented as a ribbon and the nucleobases as protruding sticks (colors are used to distinguish between different nucleotides).}
\label{fig:1}
\end{figure*}

Among these requisites, one of the biggest challenges is the realization of single-base resolution. Sufficiently thin nanoelectrodes are required, in order to have no more than one nucleotide within a close interaction range to the transverse electrodes at any given time when the nucleotides on the target DNA are passing one by one through the nanopore (\ref{fig:1}a). Considering that a nucleotide possesses dimensions of roughly 1 nm, it can be comprehended why it is extremely challenging to prepare and electrically connect sufficiently thin nanopore-embedded electrodes to achieve single-base resolution.

To potentially solve this issue, an intriguing proposal was recently made, namely to prepare a nanogap in {\it graphene} and use its edges as electrodes for DNA sequencing \cite{Postma2010}. Being a one-atom-thick planar sheet of carbon atoms \cite{Novoselov2004,CastroNeto2009}, graphene represents the ultimate limit of how thin a nanoelectrode could possibly be, and hence, the associated prospects for single-base resolution are expected to be optimal. Taking several other advantages of graphene into account, such as its ability to be tailored by nanolithography \cite{Tapaszto2008}, graphene definitely shows great promise for use as nanoelectrodes in DNA sequencing. An experimental setup related to that proposed in Ref.\ \citenum{Postma2010}, however not consisting of a nanogap, but rather a nanopore, drilled with an electron beam into graphene, was actually very recently realized independently by three research groups \cite{Dekker2010,Drndic2010,Golovchenko2010} and DNA translocation through these fabricated graphene nanopores was successfully demonstrated. In addition, a theoretical study based on density functional theory has been performed to explore the potential detection capabilities of nucleotides inside a graphene nanopore \cite{Prezhdo2010}. These achievements represent an important milestone on the road towards realization of solid-state nanopore-based DNA sequencing, and it can be expected that further advances will allow the processing of graphene to fabricate electrodes for the performance of transverse conductance measurements (as opposed to ionic blockage current measurements).

A drawback of graphene electrodes is that their conductance is much reduced compared to that of gold nanoelectrodes. This can be ascribed to several physical mechanisms: first, as a semi-metal or zero-gap semiconductor, the intrinsic electronic conductibility of graphene is smaller than that of the metal gold. Second, the coupling between graphene electrodes and DNA is smaller than that between gold nanoelectrodes and DNA due to the much smaller space extension of carbon outer orbitals compared to those of gold. Since the coupling strength drops exponentially with decreasing overlap, the transverse tunneling conductance using graphene electrodes will be reduced dramatically relative to that using gold nanoelectrodes. Such a view is also supported directly from our studies when evaluating the coupling elements which are on an average found to be larger for gold than for carbon at a given nanopore diameter. This small conductance, in turn, will lead to a deteriorated signal-to-noise ratio, i.e., the noise caused by ionic currents and by structure fluctuations of DNA during the translocation process \cite{Krems2009} will smoothen out any characteristic electrical signatures of the nucleotides, thus making it very difficult or even impossible to accurately determine the DNA sequence.

Given that the introduction of a hydrogen bond (H-bond) can enhance electron tunneling rates over vacuum tunneling \cite{Wuttke1992}, we propose here hydrogenation of the edges of graphene electrodes in order to improve their sensitivity for the DNA sequencing purpose. A somewhat weakened form of H-bonds is expected to form between the hydrogen atoms at the graphene electrode edges and those atoms carrying a partial negative charge on the DNA nucleobases, since graphene is only slightly electronegative and as a result the hydrogen atoms on the graphene edge will only carry a relatively small positive charge (calculated by us from density functional theory to be around +0.16 $e$) compared to the typical positive charge in an H-bond. For simplicity, we will continue to refer in the following to these bonds as H-bonds, but it should be understood that they are generally weaker than common H-bonds.

Several advantages can be expected from using edge-hydrogenation:
\begin{enumerate}
\item {H-bonds formed between graphene electrodes and the translocating DNA bases can enhance the coupling between them, and thus substantially increase the magnitudes of transverse tunneling currents. As a consequence, the current measurability is greatly improved and the speed with which the DNA sequence can be read is raised since the increased electrical currents no longer require a time-consuming femto-ampere amplification setup \cite{Branton2008,Zwolak2008}.}

\item {The nucleotides of the DNA molecule possess many internal degrees of freedom and can assume a large number of conformations when passing through the nanopore. These atomic conformations are overall very similar but exhibit atomic-scale differences. Considering that tunneling currents are exponentially sensitive even to atomic-scale changes of orientation and distance, severe variance of the measured conductance can be expected and hence the signal-to-noise ratio is deteriorating. The introduction of H-bonds can favorably affect the orientation and position of nucleotides when DNA is passing through the nanopore, and thus help to reduce the conductance variance.}

\item {The H-bonds will cause an attractive force which is stronger than that from van der Waals interaction, so that they can slow down the translocation of DNA through the nanopore, providing more time for the transverse conductance measurement of each nucleotide located within the nanogap between transverse electrodes, thus sampling over inevitable noise and molecular motion. There is no risk that the DNA molecule would get stuck in the nanopore due to the H-bonds because they are sufficiently weak so that they can be broken easily by the longitudinally oriented electric driving field.}
\end{enumerate}

\ref{fig:1}a gives a schematic view of solid-state nanopore-based DNA sequencing: a single-stranded DNA molecule is driven through the nanopore electrophoretically by a longitudinal electrical field while the transverse tunneling conductance is recorded for the purpose of identifying the type of the individual nucleotides. Here, the thickness of nanoelectrodes ($W_x$) should be no greater than a critical value to achieve single-base resolution of the target DNA. At the same time, it should also provide adequate coupling to the target DNA in order to obtain a sufficiently large transverse tunneling conductance. \ref{fig:1}b and \ref{fig:1}c provide cross-section views of the nanopore setup with edge-hydrogenated graphene electrodes employed in our simulations.

Our theoretical calculations and simulations were implemented as follows: first, the electrical static potential charge distribution on the graphene electrodes is determined self-consistently with density functional theory method (here, the generalized gradient approximations is employed as implemented in the BLYP exchange correlation functional). Next, taking into account the previously determined charge distribution on the graphene electrodes, translocation of single-stranded DNA through the nanopore is simulated with molecular dynamics for which the software codes NAMD2 \cite{Phillips2005} and VMD \cite{Humphrey1996} have been used following the procedure described in Ref.\ \citenum{Aksimentiev2009}. The real-time electronic structure of a translocating DNA molecule is then obtained within the extended H\"uckel model. Finally, the transverse tunneling conductivity is calculated using the Landauer-B\"uttiker formula and non-equilibrium Green's function method. (For further details about the molecular dynamics simulation settings and electrical property calculations, we refer the reader to our previous work \cite{He2010}.)

\begin{figure}[h!]
\includegraphics[width=8cm]{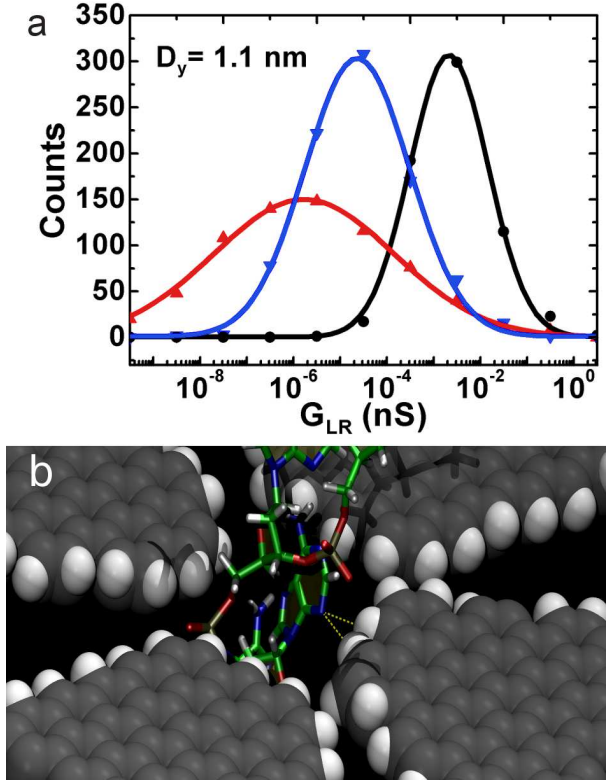}
\caption{(a) Transverse differential conductance ($G_d$) distribution curves of poly(dA)$_{30}$ translocating through a nanopore using edge-hydrogenated graphene electrodes (black line), unhydrogenated electrodes (red line), and \textit{pseudo-hydrogenated} electrodes (blue line). The nanopore inner diameter ($D_y$) amounts to 1.1 nm, the longitudinal driving field has a strength of $E_x$ = 5 kcal/(mol \AA), and the transverse bias voltage $V_{0}$ is set at 3.2 V, which is near the position of a characteristic eigen-level of adenine. (b) A snapshot extracted from the molecular dynamics simulation of the DNA translocation through the nanopore, showing a moment when two H-bonds (dotted yellow lines) are formed simultaneously between the nitrogen atom of a a DNA nucleobase and two H atoms attached to the graphene-edge. For the sake of clarity, only relevant atoms from the edge-hydrogenated graphene electrodes and the DNA molecule have been visualized, omitting water molecules, counter ions, and the silicon-nitride membrane.}
\label{fig:2}
\end{figure}

In \ref{fig:2}a we plot the transverse differential conductance distribution curves ($G_{d}$) of a poly(dA)$_{30}$ chain translocating through a nanopore with a distance of $D_{y}$ = 1.1 nm between opposing graphene electrodes for three different scenarios: edge-hydrogenated graphene electrodes (black line), unhydrogenated electrodes (red line), and \textit{pseudo-hydrogenated} electrodes (blue line). The definition of the \textit{pseudo-hydrogenated} electrodes setup will be provided below. In this work, transverse \textit{differential conductance} is employed for gathering sequencing data, because it can directly exhibit the characteristic local electronic densities of states of the target molecules and thus optimize the contrast between the different nucleotides \cite{He2010}. Here, the conductance distribution curves reveal an expansive distribution that extends over more than 3 orders of magnitude. This spread is attributed to the variation of molecule-electrode contact distances associated with the diverse molecular conformations during the translocation process through the nanopore. However, a well-defined single maximum is discernible in each conductance distribution. The position of these respective maxima indicates the conductivity of the most probable set of resembling nucleotide conformations when passing through the nanoelectrode gaps under the influence of a driving field. The width of the peak characterizes the degree of variation in orientation and position of the DNA bases when they are located between the nanoelectrodes: the sharper the peaks, the less variation in orientation and position. An extremely idealized case could be imagined where DNA translocates through the nanopore like a stiff rod with no internal degrees of freedom at all. This hypothetical scenario would correspond to a very sharp peak in the conductance distribution curve, i.e., the overlap between conductance distribution curves for different types of nucleotides would be virtually zero, and as a consequence, the corresponding electrical signatures would become fully distinguishable and the identification of individual nucleotides would be straightforward. In reality, the best-case scenario is a sufficiently narrow distribution of the measured conductance with minimal overlap between the distribution curves of different nucleotides.

In order to see the qualitative and quantitative advantages in the performance of hydrogenated graphene electrodes more directly, we have also plotted for comparison the conductance distribution curve for graphene electrodes with bare edges (i.e., unhydrogenated) in \ref{fig:2}a (red line). It can be clearly seen that hydrogenation of the edges substantially increases the transverse tunneling conductance by about 3 orders of magnitude and leads to a much more narrow distribution.

The following question naturally arises then: since, taken by itself, the introduction of hydrogen atoms shortens the atomic-scale distance between graphene electrodes and the translocating DNA molecule, resulting in an increase of the molecule-contact coupling strength and the associated transverse tunneling current, how much contribution to the increased conductance originates from the actual formation of H-bonds (as shown in \ref{fig:2}b), and how much is due to the mere presence of hydrogen atoms at the contact? In order to quantitatively answer this question we eliminate the contribution of the H-bonds while preserving the orbital overlap by replacing the hydrogen atoms on the graphene-edge by a set of  dummy hydrogen-like atomic orbitals in our simulations. In this artificial scenario, which we refer to as \textit{pseudo-hydrogenated}, no H-bonds can be formed at the contact and any increase of the transverse tunneling conductance originates only from the shortened molecule-contact distance. The obtained conductance distribution curve is plotted in \ref{fig:2}a (blue line). By comparing the peak positions between the three cases, it becomes apparent that the H-bond formed at the contact is responsible for the most substantial contribution (about 2 orders of magnitude) to the increase in the transverse tunneling conductance. Furthermore, by comparing the peak widths of the three cases, we find that \textit{both} the hydrogenation and the pseudo-hydrogenation of the graphene electrode edges lead to a significant reduction in the conductance variation. It thus appears that both confining effect from the hydrogen atoms at the graphene edge, and orientation effects caused by the formation of H-bonds play a role. Quantitative evaluation of the confining effect is achieved by comparing the peak width of the conductance distribution curve obtained using pseudo-hydrogenated graphene and that using unhydrogenated graphene, while an evaluation of orientation effects can be achieved by comparing the cases of edge-hydrogenated graphene and of pseudo-hydrogenated graphene. This analysis shows that for nanopores with very small diameters, confining effects play the major role for the reduction of the conductance variation.

\begin{figure}[ht]
\includegraphics[width=8cm]{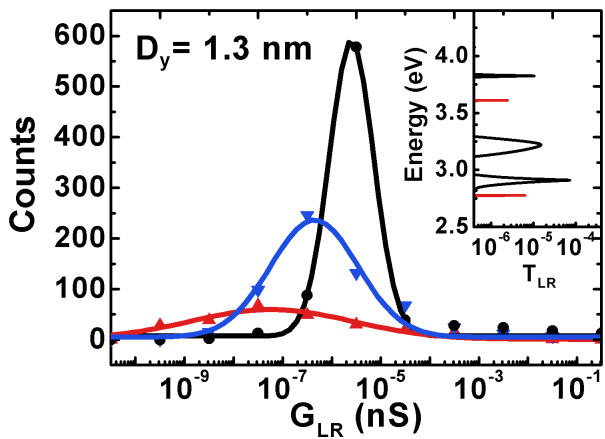}
\caption{Transverse differential conductance ($G_d$) distribution curves of poly(dA)$_{30}$ translocating through a nanopore using edge-hydrogenated graphene electrodes (black line), unhydrogenated electrodes (red line), and \textit{pseudo-hydrogenated} electrodes (blue line). The nanopore inner diameter ($D_y$) is 1.3 nm wide, the longitudinal driving field has a strength of $E_x$ = 5 kcal/(mol \AA), and the transverse bias voltage $V_{0}$ is set at 3.2 V. The inset shows the transverse transmission spectra at a random snapshot during the translocation process for edge-hydrogenated graphene electrodes (black line) and for unhydrogenated graphene electrodes (red line).}
\label{fig:3}
\end{figure}

In order to maximize the chance for the formation of H-bonds between edge-hydrogen atoms on the graphene electrodes and the translocating DNA bases, the gap between the nanoelectrodes ($D_{y}$) should be made as narrow as possible. To explore the dependence of H-bond formation on the nanopore inner diameter, we simulated the translocation of poly(dA)$_{30}$ through a nanopore with $D_{y}$ = 1.3 nm, which is just about one row of carbon atoms in graphene wider than the 1.1 nm of the previously considered case presented in \ref{fig:2}. The calculated conductance distribution curves are plotted in \ref{fig:3}, where the data for the system using edge-hydrogenated graphene electrodes is drawn in black, that using unhydrogenated electrodes in red, and that using \textit{pseudo-hydrogenated} electrodes in blue. 

\ref{fig:3} clearly demonstrates that although the overall conductivity is significantly reduced when compared to the system with $D_{y}$ = 1.1 nm (an unavoidable side effect of the larger nanoelectrode gap), edge-hydrogenation of graphene leads to an even more prominent improvement of conductance measurability and reduction of conductance variance: when using unhydrogenated graphene electrodes, the transverse tunneling conductance is too small (about $10^{-8}$ nS) and possesses a too broad distribution (about 5 orders of magnitude) to be of any use for DNA nucleotide detection, while hydrogenation of graphene electrodes can significantly improve the conductivity (by about 3 orders of magnitude) and reduce the associated variance (amounting to merely about 2 orders of magnitude). 

Upon comparison with the artificial scenario of using \textit{pseudo-hydrogenated} electrodes, it becomes apparent that the confining effect and the orientation effect play nearly equal roles for the reduction of conductance variance in case of a larger nanopore diameter. The physical mechanism of these substantial changes can be identified from the inset of \ref{fig:3}, where the transverse transmission spectra at a random snapshot are plotted for poly(dA)$_{30}$ translocation with edge-hydrogenated graphene electrodes (black line) and with unhydrogenated graphene electrodes (red line): edge-hydrogenation causes the transmission peaks to become higher and wider, indicating much better coupling of the DNA molecule with the transverse electrodes.

Another striking observation from the plots in \ref{fig:2,fig:3} is that the sum of counts under the distribution curves is significantly increased when considering edge-hydrogenation on the graphene electrodes. Our analysis shows that two factors contribute to this increase: one is that for transverse conductance measurements using unhydrogenated graphene electrodes, quite a few of the results are below the threshold of $10^{-10}$ nS to be realistically measurable in any experiments, and are therefore dropped from the distribution curve; another is that by using edge-hydrogenated graphene electrodes the DNA translocation speed is reduced. Reducing the DNA translocation speed during the detection process is a crucial requisite for DNA base identification in nanopores, since each nucleotide should remain between the transverse nanoelectrodes sufficiently long to sample over any unavoidable noise background. This slower translocation speed can also be directly observed in our molecular dynamics simulations, which exhibit a reduction of about 20\% in speed when H-bonds are formed. For a weaker electric driving field, this percentage-wise reduction could be even higher.

But it can also be noted in \ref{fig:2} that the number of counts for the 1.1 nm wide nanopore is about the same for both hydrogenated and pseudo-hydrogenated case. Therefore, the formation of H-bonds alone cannot fully explain our results. Rather, the hindrance of DNA translocation through the nanopore for smaller diameters by repulsive interaction due to presence of graphene-edge hydrogen atoms is also responsible for a slower translocation speed (as observed in our molecular dynamics simulations). For the larger nanopore diameter of 1.3 nm (\ref{fig:2}), this hindrance becomes less of an issue, as the DNA molecule has sufficient space to evade the protruding hydrogen atoms at the graphene edges. In that case, the formation of H-bonds becomes the main effect for the slowing-down of the translocation process, as can be seen from the larger number of counts for the hydrogenated over the pseudo-hydrogenated case.

\begin{figure}[ht]
\includegraphics[width=8cm]{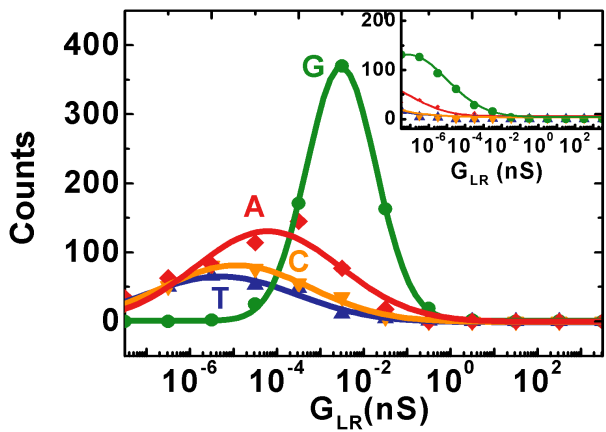}
\caption{Transverse differential conductance ($G_d$) distribution curves of poly(dX)$_{30}$ (X = A, T, C, G) translocating through a nanopore using edge-hydrogenated graphene electrodes. The gap between the transverse electrodes ($D_y$) is 1.1 nm, the longitudinal driving field $E_x$ = 5 kcal/(mol \AA), and the transverse bias voltage $V_{0}$ is set at 4.8 V, near the position of a LDOS maximum of guanine. The inset shows the corresponding results obtained using bare-edge (unhydrogenated) graphene electrodes.}
\label{fig:4}
\end{figure}

\ref{fig:4} plots $G_{d}$ distribution curves of poly(dX)$_{30}$ (where X stands for A, T, C, and G, respectively) translocating through a nanopore with $D_{y}$ = 1.1 nm. Data for edge-hydrogenated graphene electrodes is shown in the main panel of the figure, while data for the unhydrogenated system is plotted in the inset. Here, the transverse bias voltage $V_{0}$ is set at 4.2 V where the local electronic density of states (LDOS) for guanine has a maximum. As a result, $G_{d}$ of G is orders of magnitude larger than that of the other three nucleotides, making it possible to easily identify G. Although it appears from \ref{fig:4} as if the bases A, C, and T could not be distinguished from another, one should keep in mind that a proper adjustment of $V_{0}$ to the positions of the respective maxima in the LDOS of other nucleotides can actually resolve these differences, as we have shown recently on the example of a setup based on gold nanoelectrodes \cite{He2010}.

Finally, it should be emphasized that, according to our calculations, the likelihood for H-bond formations becomes dramatically enhanced when the nanopore inner diameter $D_{y}$ is about 1.3 nm or less. This requirement for the nanopore diameter originates from the comparatively small partial positive charges on the graphene-edge hydrogen atoms: considering that the hydrogen atoms carry a charge of about +0.16 $e$ (calculated within the framework of density functional theory) and not the approximately +0.35 to +0.45 $e$ typical for hydrogen atoms forming usual H-bonds, the distances at which DNA nucleobases are effectively attracted towards the graphene edge to form these weakened H-bonds should be rather small. Experimentally, the preparation of such tiny nanopores is expected to be extremely challenging. However, we would like to point out that our analysis of the effect of H-bonds on the transverse tunneling conductance is in principle not limited to the hydrogenation of graphene edges. In fact, H-bonds could be introduced in other ways, such as, e.g., by attaching a functional group to the nanoelectrodes. Previous theoretical calculations \cite{He2008} indicate that a nanoelectrode chemically functionalized with a probing base can form H-bonds with the DNA nucleotides as well. Our analysis and conclusions still apply, and such H-bond-assisted nucleotide recognition has in fact been verified experimentally \cite{Ohshiro2006,Chang2009,Chang2010}.

In summary, through edge-hydrogenation of the graphene electrodes in a nanopore-based DNA sequencing setup, the transverse tunneling conductance can be drastically raised by about 3 orders of magnitude, thus improving the conductance measurability substantially. At the same time, the variation in the conductance will be significantly reduced, leading to a faster and more reliable identification of the four nucleotide types. The pico-siemens tunneling conductance facilitates reading the nucleotide sequence at a much greater speed than what is possible with only femto-ampere tunneling currents. The analysis of our simulation shows that the formation of H-bonds between the hydrogenated graphene edges and the DNA nucleobases plays a crucial role in the described improvements.

%% classes.  Rather than use \section, an appropriate macro is
%% provided that will always work.
%%%%%%%%%%%%%%%%%%%%%%%%%%%%%%%%%%%%%%%%%%%%%%%%%%%%%%%%%%%%%%%%%%%%%
\acknowledgement

We gratefully acknowledge financial support from the National Science Fund for Distinguished Young Scholars (Grant No.\ 60825403), China Ministry of Science and Technology (Contract No.\ 2010CB934200), the Swedish Foundation for International Cooperation in Research and Higher Education (STINT), the Swedish Research Council (VR, Grant No.\ 621-2009-3628), Wenner-Gren Foundations, Carl Tryggers Stiftelse f\"or Vetenskaplig Forskning, and the Uppsala University UniMolecular Electronics Center (U$^3$MEC).

%% \suppinfo macro.
%%%%%%%%%%%%%%%%%%%%%%%%%%%%%%%%%%%%%%%%%%%%%%%%%%%%%%%%%%%%%%%%%%%%%
%\suppinfo

\bibliography{achemso}

\end{document}